\documentclass[12pt]{iopart}

\usepackage{iopams}
\usepackage[dvipdfmx]{graphicx}

\usepackage{xcolor}
\expandafter\let\csname equation*\endcsname\relax
\expandafter\let\csname endequation*\endcsname\relax
\usepackage{amsmath}
\usepackage{bm}
\usepackage{braket}
\newcommand{\new}{\nonumber\\}

\newcommand{\tx}{\tilde{x}}
\newcommand{\tp}{\tilde{p}}
\newcommand{\tV}{\tilde{V}}
\newcommand{\tb}{\tilde{\beta}}

\newcommand{\abs}[1]{\left|#1\right|}
\newcommand{\ave}[1]{\left\langle #1\right\rangle}

\begin{document}

\title{ Quantum fluctuations can enhance or reduce positional
uncertainty at finite temperature }

\author{Harukuni Ikeda$^{1}$}

\address{
$^1$Yukawa Institute for Theoretical Physics, Kyoto University,
Kyoto 606-8502, Japan}
\ead{harukuni.ikeda@yukawa.kyoto-u.ac.jp}
\vspace{10pt}
\begin{indented}
\item[]\today 
\end{indented}

\begin{abstract}
The uncertainty principle guarantees a non-zero value for the positional
uncertainty, $\ave{\Delta x^2} > 0$, even without thermal
fluctuations. This implies that quantum fluctuations inherently enhance
positional uncertainty at zero temperature. A natural question then
arises: what happens at finite temperatures, where the interplay between
quantum and thermal fluctuations may give rise to complex and intriguing
behaviors? To address this question, we systematically investigate the
positional uncertainty, $\ave{\Delta x^2}$, of a particle in equilibrium
confined within a nonlinear potential of the form $V(x) \propto x^n$,
where $n = 2, 4, 6, \dots$ represents an even exponent. Using path
integral Monte Carlo simulations, we calculate $\ave{\Delta x^2}$ in
equilibrium as a function of the thermal de Broglie wavelength
$\Lambda$. Interestingly, for large values of $n$, $\ave{\Delta x^2}$
exhibits a non-monotonic dependence on $\Lambda$: it initially decreases
with increasing $\Lambda$ at small $\Lambda$ but increases at larger
$\Lambda$. To further understand this behavior, we employ a
semiclassical approximation, which reveals that quantum fluctuations can
reduce positional uncertainty for small $\Lambda$ when the nonlinearity
of the potential is sufficiently strong. Finally, we discuss the
potential implications of this result for many-body phenomena driven by
strong nonlinear interactions, such as glass transitions, where the
transition densities exhibit a similar non-monotonic dependence on
$\Lambda$.
\end{abstract}

%\tableofcontents  

%\newpage
\section{Introduction}
In a classical system, the position is a deterministic variable in the
absence of thermal fluctuations, {\it i.e.}, the positional uncertainty
is zero, $\ave{\Delta x^2} \equiv \ave{x^2}-\ave{x}^2= 0$. In contrast,
for a quantum system, the uncertainty principle ensures $\ave{\Delta
x^2} > 0$~\cite{greiner2011quantum}, indicating that quantum
fluctuations inherently increase positional uncertainty. What happens at
finite temperatures in equilibrium? 
This study systematically
investigates the effects of quantum fluctuations on $\ave{\Delta x^2}$
in equilibrium using both numerical and theoretical approaches. To
measure the relative strength of quantum fluctuations, we use the
thermal de Broglie wavelength in units of the linear size of the system $L$:
\begin{align}
 \Lambda = \sqrt{\frac{\beta\hbar^2}{mL^2}},
\end{align}
where $\hbar$ denotes the reduced Planck constant, $\beta$ denotes the
inverse temperature, $m$ denotes the mass of the particle, and $L$
denotes the linear size of the system.
In this work, we investigate how the system behaves 
when $\Lambda$ is systematically varied by changing $m$ or $\hbar$, 
while keeping $\beta$ fixed.

A naive intuition suggests that, at a fixed temperature, $\ave{\Delta
x^2}$ should increase with increasing $\Lambda$, owing to tunneling
effects and zero-point energy~\cite{greiner2011quantum}. This intuition
holds true in the simplest case: a particle confined in a harmonic
potential, $V(x) \propto x^2$, where $\ave{\Delta x^2}$ increases
monotonically as $\Lambda$
increases~\cite{greiner2012thermodynamics}. However, this behavior does
not generalize to all systems, as seen in another analytically solvable
example: a particle confined between hard walls. For a classical
particle at finite temperature, the positional distribution is uniform
between the walls. In contrast, for finite $\Lambda$, quantum effects
become significant. The boundary conditions and the continuity of the
wave function suppress the probability of finding the particle near the
walls. As a result, $\ave{\Delta x^2}$ becomes smaller than its
classical counterpart ($\Lambda = 0$), demonstrating that quantum
fluctuations can suppress positional uncertainty.

The observation that quantum fluctuations can suppress positional
uncertainty is rather counterintuitive. Interestingly, similar behaviors
have been reported in many-body quantum systems. For example, quantum
fluctuations often reduce the uncertainty in the rotational degrees of
freedom of hydrogen molecules in solid phases, which stabilizes the
so-called phase II (the partially frozen
phase)~\cite{kitamura2000quantum,hemley2000element,tsuneyuki2002quantum}.
Related phenomena have also been observed in
crystallization~\cite{hansen1971,runge1988,sese2007computational,sese2007computationalII,sese2013path,yamashita2014gas},
glass
transitions~\cite{markland2011,zamponi2011,markland2012,biroli2012tentative,kinugawa2021,das2021,das2022,winer2024},
and spin-glass
transitions~\cite{foini2010,foini2011,thomson2020,urbani2024}, where
quantum fluctuations reduce the transition densities. An intuitive
explanation for these phenomena is that quantum fluctuations effectively
increase the particles’ radii, thereby reducing the accessible volume
and lowering the transition
density~\cite{markland2011,zamponi2011}. This situation is analogous to
the case of a single particle confined between hard walls, where quantum
fluctuations, as described earlier, prevent the particle from occupying
regions near the walls. This effectively reduces the accessible volume.

An important and natural question is: under what conditions do quantum
fluctuations suppress positional uncertainty? To address this question,
we systematically study a particle confined in a nonlinear
potential, $V(x) \propto x^n$, where $n = 2, 4, \dots$ represents an
even number. By varying $n$, we can systematically change the functional
form from harmonic ($n=2$) to hard walls ($n\to\infty$).

We investigate the model for $2 < n < \infty$ using path-integral Monte
Carlo simulations~\cite{feynman2010quantum,tuckerman1993,ceperley1995,
habershon2013ring}. For weakly nonlinear potentials ({\it i.e.}, small $n$),
$\ave{\Delta x^2}$ increases monotonically with increasing $\Lambda$,
similar to the behavior observed in the harmonic potential
case. Interestingly, for large $n$, $\ave{\Delta x^2}$ decreases from
its classical value at moderate $\Lambda$, indicating that quantum
fluctuations suppress the positional uncertainty. This effect becomes
most pronounced in the hard-wall limit ($n \to \infty$), where
$\ave{\Delta x^2}$ decreases monotonically with $\Lambda$. To further
understand this behavior, we perform a semi-classical
calculation~\cite{feynman2010quantum,zinn2021quantum} to determine the
conditions under which quantum fluctuations suppress $\ave{\Delta
x^2}$. The semi-classical analysis predicts that $\ave{\Delta x^2}$
increases for $n = 2, 4$ and decreases for $n > 4$ at small $\Lambda$,
consistent with the numerical results. These findings demonstrate that
quantum fluctuations can generally suppress positional uncertainty at
small $\Lambda$ when the potential is strongly nonlinear.

The structure of the paper is as follows. In Sec.\ref{172210_11Nov24},
we introduce the model. In Sec.\ref{111853_17Jan25}, we briefly review
two analytically solvable cases: the harmonic potential and hard
walls. In Sec.\ref{172251_11Nov24}, we present the numerical results
obtained through path-integral Monte Carlo simulations. In
Sec.\ref{172319_11Nov24}, we provide analytical insights derived from
the semi-classical calculation. Finally, Sec.~\ref{172700_11Nov24} is
dedicated to the summary and discussions, highlighting some potential
implications of our findings for studies of glass transitions.

\section{Settings}
\label{172210_11Nov24} In this section, we introduce the model and
define key physical quantities.

\subsection{Model}
\begin{figure}[t]
\begin{center}
\includegraphics[width=10cm]{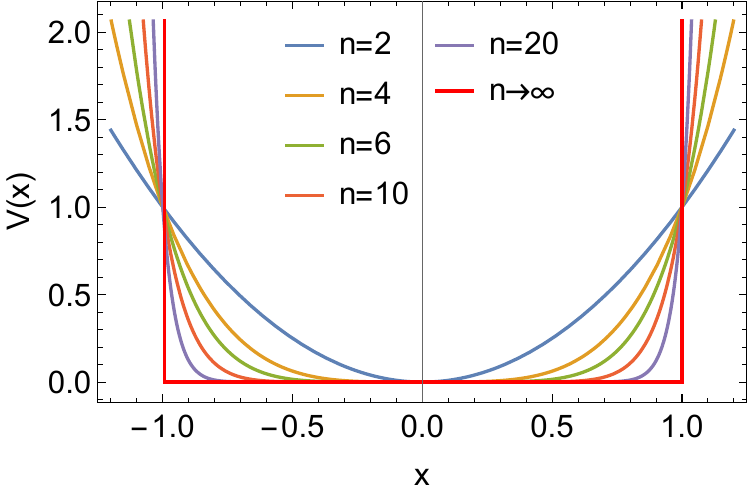} \caption{Potentials for several
values of $n$. The potential is harmonic when $n = 2$, and it becomes
equivalent to hard walls in the limit $n \to \infty$. For simplicity, we
set $k = 1$ and $L = 1$ in this figure.}  \label{150350_26Oct24}
\end{center}
\end{figure}

We consider a particle in one dimension confined within a potential
$V(x)$. The Hamiltonian of the system is given by:
\begin{align}
H = \frac{p^2}{2m} + V(x),
\end{align}
where the potential $V(x)$ is defined as:
\begin{align}
V(x) = k\left(\frac{x}{L}\right)^n, \label{144742_26Oct24}
\end{align}
with $n = 2, 4, 6, \dots$ representing an even number. The position $x$
and momentum $p$ satisfy the canonical commutation
relation~\cite{greiner2011quantum}:
\begin{align}
[x, p] = i\hbar.
\end{align}
For $n = 2$, $V(x)$ corresponds to a harmonic potential, while in the
limit $n \to \infty$, it approaches the case of hard walls, as shown in
Fig.~\ref{150350_26Oct24}.

To quantify the positional uncertainty, we observe the variance of $x$
in equilibrium:
\begin{align}
&\ave{\Delta x^2} \equiv \frac{1}{Z} \tr \left(e^{-\beta H}x^2\right),
&Z = \tr e^{-\beta H},
\end{align}
where $\beta = 1 / (k_B T)$ is the inverse temperature, and $k_B$ is the
Boltzmann constant.

\subsection{Nondimensionalization}
We introduce dimensionless variables:
\begin{align}
&\tx = L^{-1}x,
&\tp = \frac{L}{\hbar}p,
\end{align}
which satisfy the commutation relation:
\begin{align}
[\tx, \tp] = i.
\end{align}
With these dimensionless quantities, the partition function can be
rewritten as:
\begin{align}
Z = \tr \exp\left[-\Lambda^2\frac{\tp^2}{2} - \tb\tV(\tx)\right], \label{113623_12Nov24}
\end{align}
where:
\begin{align}
&\Lambda = \sqrt{\frac{\beta\hbar^2}{mL^2}}, \
&\tb = k\beta, \
&\tV(\tx) = \abs{\tx}^n.
\end{align}
After nondimensionalization, the system is governed by two control
parameters: the reduced inverse temperature $\tb$ and the thermal de
Broglie wavelength $\Lambda$~\cite{greiner2012thermodynamics}.

\section{Solvable Cases}
\label{111853_17Jan25}

Here we shortly revisit the well known solvable cases.

\subsection{Analytical Results for $n=2$ (Harmonic Potential)}
\begin{figure}[t]
\begin{center}
\includegraphics[width=10cm]{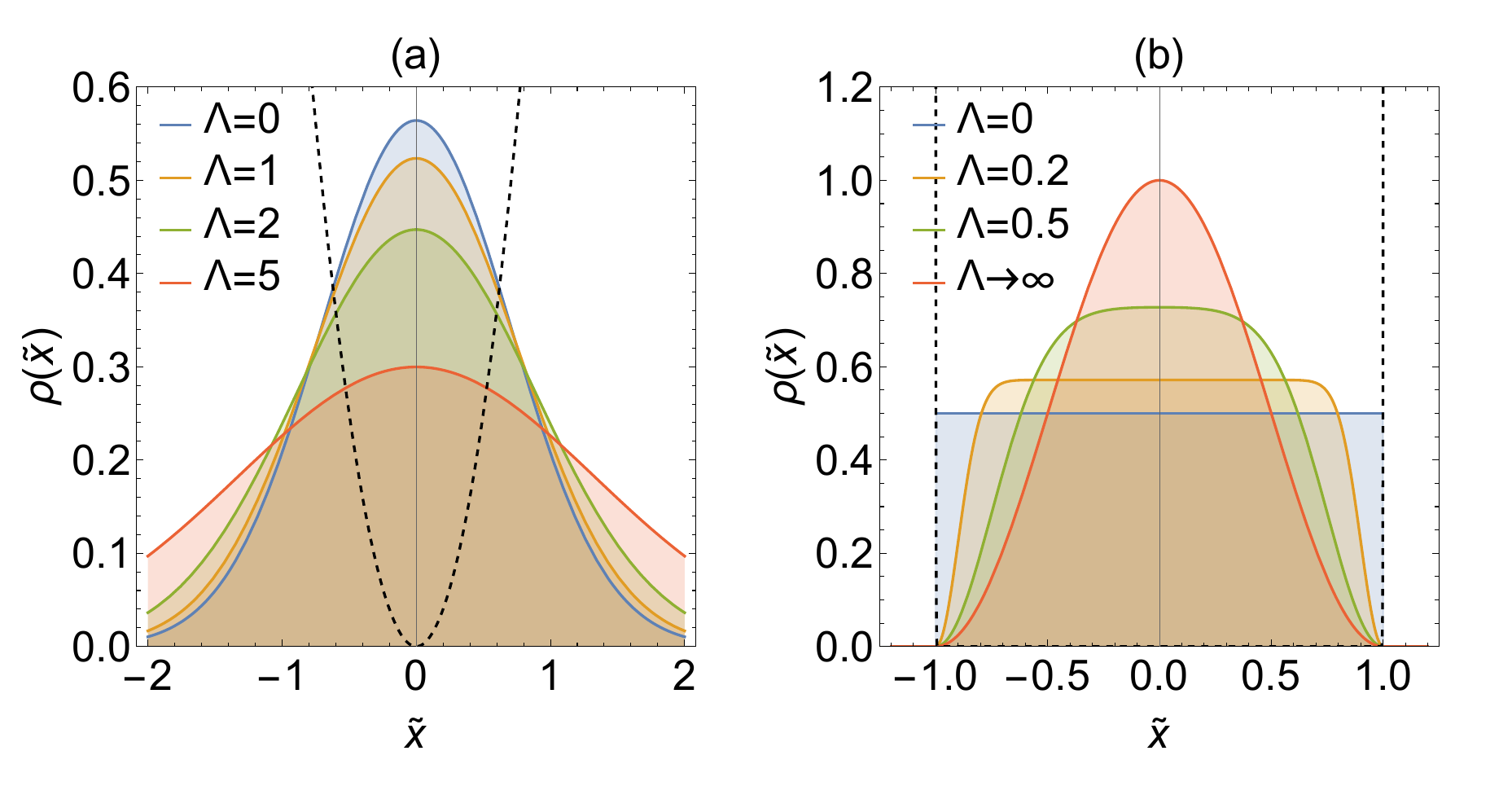} \caption{Equilibrium
distributions $\rho(\tx)$ for $\tb=1$. (a) $\rho(\tx)$ for $n=2$
(harmonic potential). Solid lines represent $\rho(\tx)$ for various
values of the de Broglie wavelengths
$\Lambda=\sqrt{\frac{\beta\hbar^2}{mL^2}}$, while the black dashed line
represents the potential $V(\tx)=\tx^2$. The distribution $\rho(\tx)$
widens as $\Lambda$ increases. (b) Same plots for the limit $n\to\infty$
(hard walls). In this case, the distribution becomes narrower as
$\Lambda$ increases.}  \label{134304_3Nov24}
\end{center}
\end{figure}

The model can be solved analytically for $n=2$, where $V(x)$ is a
harmonic potential. In this case, the equilibrium distribution of $\tx$
follows a Gaussian
distribution~\cite{greiner2012thermodynamics,schnhammer2014}:
\begin{align}
\rho(\tx) = \frac{1}{Z} \braket{\tx | e^{-\beta H} | \tx}
= \frac{1}{\sqrt{2\pi\ave{\Delta\tx^2}}}
e^{-\frac{\tx^2}{2\ave{\Delta\tx^2}}},
\end{align}
where the variance is given by:
\begin{align}
\ave{\Delta\tx^2} = \frac{\Lambda}{2\sqrt{2\tb}}
\coth\left(\Lambda\sqrt{\frac{\tb}{2}}\right).
\end{align}
As shown in Fig.~\ref{134304_3Nov24}~(a), the distribution $\rho(\tx)$
broadens with increasing $\Lambda$, indicating that quantum fluctuations
enhance the positional uncertainty. This result is expected, as
tunneling effects and zero-point energy weaken the confinement imposed
by the potential.
In Fig.~\ref{100335_29Oct24}, we plot $\ave{\Delta\tx^2}$ as the blue
solid line, showing its monotonic increase with $\Lambda$.

\subsection{Analytical Results for $n\to\infty$ (Hard Walls)}
In the limit $n\to\infty$, the model corresponds to a particle confined
between hard walls at $\tx=\pm 1$. For the classical case $\Lambda=0$,
the equilibrium distribution is flat:
\begin{align}
\lim_{\Lambda\to 0}\rho(\tx) =
\begin{cases}
1/2 & \abs{\tx} < 1, \\
0 & \abs{\tx} \geq 1,
\end{cases}
\end{align}
which results in a positional uncertainty:
\begin{align}
\ave{\Delta \tx^2}_0 =\frac{1}{3}.
\end{align}
For $\Lambda > 0$, the equilibrium distribution is expressed as:
\begin{align}
&\rho(\tx) = \frac{1}{Z} \sum_{n=1}^\infty e^{-\frac{(n\pi\Lambda)^2}{8}} 
\abs{\phi_n(\tx)}^2,
&Z = \sum_{n=1}^\infty e^{-\frac{(n\pi\Lambda)^2}{8}},
\end{align}
where $\phi_n(\tx)$ denotes the $n$-th eigenfunction:
\begin{align}
\phi_n(\tx) = \sin\left(n\pi \frac{\tx-1}{2}\right). \label{135010_3Nov24}
\end{align}
Due to the boundary conditions $\phi_n(\tx = \pm 1) = 0$ and the
continuity of the wave function, the eigenfunctions’ amplitudes are
significantly suppressed near the walls, $\abs{\phi_n(\tx)} \ll 1$ for
$\abs{\tx} \approx 1$. This results in a narrower distribution of
$\rho(\tx)$ compared to the classical case, as shown in
Fig.~\ref{134304_3Nov24}~(b). In the limit $\Lambda \to \infty$, the
distribution converges to that of the ground state:
\begin{align}
\lim_{\Lambda\to\infty}\rho(\tx)
 = \abs{\sin\left(\pi \frac{\tx-1}{2}\right)}^2,
\end{align}
which results in the positional uncertainty:
\begin{align}
\ave{\Delta\tx^2}_{\infty} = \frac{1}{3} - \frac{2}{\pi^2}.
\end{align}
In Fig.~\ref{100335_29Oct24}, we plot $\ave{\Delta\tx^2} = \int d\tx ,
\rho(\tx) \tx^2$ as the red solid line. The positional uncertainty
$\ave{\Delta\tx^2}$ monotonically decreases with $\Lambda$ from
$\ave{\Delta\tx^2}_0$ to $\ave{\Delta\tx^2}_{\infty}$.

\section{Numerical Simulation}
\label{172251_11Nov24}

For $2 < n < \infty$, the model cannot be solved analytically. In this
section, we present the numerical results for these cases.

\subsection{Path Integral Formulation}

To investigate the model numerically, we reformulate the partition
function $Z$ using the path-integral
formalism~\cite{feynman2010quantum,tuckerman1993,habershon2013ring,mittal2020path}:
\begin{align}
Z = \lim_{N\to\infty}\left(\frac{N}{2\pi\Lambda^2}\right)^{N/2}
\left(\prod_{i=1}^N \int d\tx_i\right)
\exp\left[-\tb \Phi(\tx_1,\cdots,\tx_N)\right], \label{224635_28Oct24}
\end{align}
where we introduce the effective potential:
\begin{align}
\Phi(\tx_1,\cdots, \tx_N) =
\sum_{i=1}^N \left[
\frac{N}{2\tb\Lambda^2} \left(\tx_{i+1}-\tx_i\right)^2+\frac{\tV(\tx_i)}{N}
\right]. \label{230808_28Oct24}
\end{align}
The thermal average of a physical quantity $A(\tx)$ can then be
expressed as~\cite{tuckerman1993}:
\begin{align}
\ave{A(\tx)} = \lim_{N\to\infty}\frac{1}{N}\sum_{i=1}^N \ave{A(\tx_i)},
\end{align}
where $\ave{\bullet}$ denotes the thermal average over the equilibrium
configurations under the effective potential $\Phi$.

\subsection{Details of Numerical Implementation}

We sample the equilibrium configurations ${\tx_1,\cdots, \tx_N}$ using
Monte Carlo (MC) simulations with the effective potential $\Phi$ in
Eq.~(\ref{230808_28Oct24}). A single MC step involves $N$ local updates
of $\tx_i \to \tx_i + \delta_x$ for a randomly chosen $i$, along with a
center-of-mass update $R \to R + \delta_R$ to accelerate
relaxation. Here, $\delta_x$ and $\delta_R$ are small random numbers
with $\delta_x \in [-0.05\Lambda, 0.05\Lambda]$ and $\delta_R \in
[-0.05, 0.05]$. Each update is accepted with probability $\min[1,
e^{-\tb\Delta \Phi}]$, where $\Delta \Phi$ represents the energy change
caused by the update. In this study, we present results for
$N=50$. Simulations with $N=100$ were also performed to confirm that the
results are independent of $N$. We equilibrated the system using $10^7$
MC steps, followed by an additional $10^7$ steps to compute the thermal
average of $\ave{\Delta\tx^2}$.

\subsection{Results}

\begin{figure}[t]
\begin{center}
\includegraphics[width=10cm]{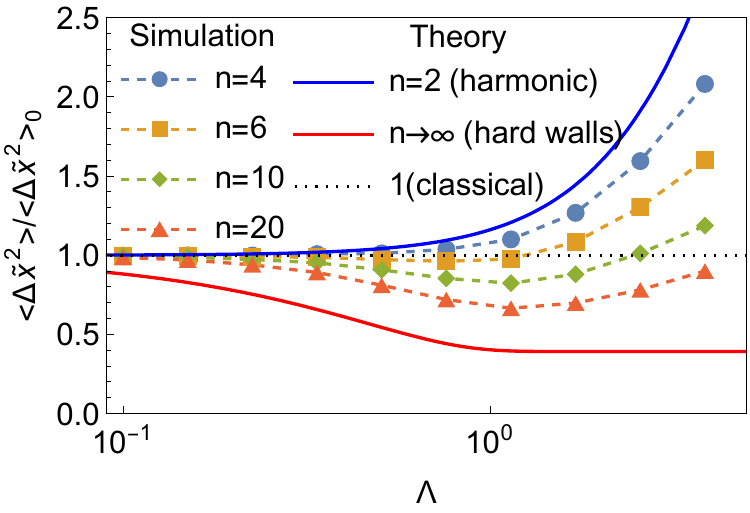} \caption{Thermal de Broglie wavelength
$\Lambda=\sqrt{\frac{\beta\hbar^2}{mL^2}}$ dependence of
$\ave{\Delta\tx^2}$, normalized by the classical result
$\ave{\Delta\tx^2}_0$. Here, we set $\tb=1$. Markers indicate numerical
results for $n=2, \dots, 20$, and solid lines represent analytical
results for $n=2$ and $n \to \infty$. The black dotted line corresponds
to $\ave{\Delta\tx^2}/\ave{\Delta\tx^2}_0 = 1$. For $n=2$ and $4$,
$\ave{\Delta\tx^2}$ increases monotonically, while for $n=6, 10$, and
$20$, $\ave{\Delta\tx^2}$ decreases for small $\Lambda$ and increases
for large $\Lambda$. See Fig.~\ref{225603_3Nov24} for an enlarged view
of the data for small $\Lambda$.} \label{100335_29Oct24}
\end{center}
\end{figure}

Fig.~\ref{100335_29Oct24} summarizes our numerical results for various
values of $n$. For small $n$ ($n=2$ and $4$), $\ave{\Delta\tx^2}$
increases monotonically with $\Lambda$, indicating that quantum
fluctuations enhance positional uncertainty. In contrast, for larger $n$
($n=6$, $10$, and $20$), $\ave{\Delta\tx^2}$ exhibits non-monotonic
behavior: it decreases for small $\Lambda$ and increases for large
$\Lambda$. These results suggest that quantum fluctuations can reduce
positional uncertainty for small $\Lambda$. Interestingly, similar
non-monotonic behaviors have been reported in previous studies of the
glass transitions~\cite{markland2011,markland2012}.

\section{Semi-classical approximation}
\label{172319_11Nov24}

We employ the semi-classical approximation to get a physical insight
into the non-monotonic behavior for large $n$ and small $\Lambda$.
Notably, the decreases of $\ave{\Delta\tx^2}$ occur for small
$\Lambda$, allowing us to determine whether $\ave{\Delta\tx^2}$
decreases or not by analyzing the sign of the leading-order correction
in the semi-classical
approximation~\cite{feynman2010quantum,biroli2012tentative}.

\subsection{Partition function}
We begin by decomposing $\tx_i$ into a center-of-mass term and fluctuations:
\begin{align}
\tx_i = R + u_i,
\end{align}
where $R = N^{-1}\sum_{i=1}^N \tx_i$ represents the center-of-mass, and
$u_i$ are fluctuations around it. For $\Lambda \ll 1$, the fluctuations
are small, $\abs{u_i} \ll 1$, enabling an expansion of $\tV(\tx_i)$
around $R$. The partition function then becomes:
\begin{align}
Z 
&\approx \lim_{N\to\infty}
 \left(\frac{N}{2\pi\Lambda^2}\right)^{N/2}
 \int dR \prod_{i=1}^N \int du_i \delta\left(N^{-1}\sum_{i}u_i\right)\new 
&\times 
e^{ -\frac{N}{2\Lambda^2}\sum_{i=1}^N(u_{i+1}-u_i)^2-\tb \tV(R)
-\frac{\tb \tV''(R)}{2N}\sum_i u_i^2}.\label{201312_20Nov24}
\end{align}
After integrating out the Gaussian fluctuations $u_i$, the partition
function reduces to~\cite{feynman2010quantum,zinn2021quantum}
\begin{align}
Z\propto \int dR
 \exp\left[-\tb \tV_{\rm eff}(R)\right],
\end{align}
where the effective potential is given by 
\begin{align}
\tV_{\rm eff}(R) = \tV(R)+\frac{\Lambda^2}{24}\tV''(R).\label{163521_3Nov24}
\end{align}
The above equation implies that the distribution of the center-of-mass
can be identified with that of a classical particle confined in the
effective potential Eq.~(\ref{163521_3Nov24}) at the level of the
semi-classical approximation~\cite{feynman2010quantum,zinn2021quantum}.

\subsection{Positional uncertainty}
The positional uncertainty can be written as 
\begin{align}
\ave{\Delta\tx^2}
= \ave{R^2} + \lim_{N\to\infty}
 \frac{1}{N}\sum_{i=1}^N\ave{u_i^2}.\label{210041_23Nov24}
\end{align}
The first term in Eq.~(\ref{210041_23Nov24}) is approximated as 
\begin{align}
&\ave{R^2} \approx
\frac{\int dR e^{-\tb \tV_{\rm eff}(R)}R^2}{\int dRe^{-\tb \tV_{\rm eff}(R)}}\new 
&\approx \ave{R^2}_{0} -\frac{\Lambda^2}{24}
 \left[\ave{\tV''(R)R^2}_{0}-\ave{\tV''(R)}_{0}
 \ave{R^2}_{0}\right],\label{170855_3Nov24}
\end{align}
where $\ave{\bullet}_{0}= \left.\ave{\bullet}\right|_{\Lambda=0}$
denotes the classical equilibrium average for $\Lambda=0$:
\begin{align}
\ave{\bullet}_{0}
 = \frac{\int dRe^{-\tb \tV(R)}\bullet }{\int dRe^{-\tb \tV(R)}}.
\end{align}
Since $\ave{\tV''(R)R^2}_{0}\geq \ave{\tV''(R)}_{0}\ave{R^2}_{0}$, the
$O(\Lambda^2)$ order term in Eq.~(\ref{170855_3Nov24}) always give
negative contribution, which reduces the positional uncertainty
$\ave{\Delta\tx^2}$. Note that this term appears only for non-linear
potentials since the $O(\Lambda^2)$ order term in
Eq.~(\ref{170855_3Nov24}) vanishes for the harmonic potential
${\tV(R)=R^2}$. The second term in Eq.~(\ref{210041_23Nov24}) is
calculated as~\cite{feynman2010quantum}
\begin{align}
\lim_{N\to\infty}\frac{1}{N}\sum_{i=1}^N \ave{u_i^2} \approx \frac{\Lambda^2}{12},\label{145524_25Nov24}
\end{align}
which gives a positive contribution and enhances the positional
uncertainty. Due to the competition of the terms in
Eq.~(\ref{170855_3Nov24}) and (\ref{145524_25Nov24}), the least-order
contribution changes the sign at a certain value of $n$, which
determines whether $\ave{\Delta\tx^2}$ increase or decrease for small
$\Lambda$.

\begin{figure}[t]
\begin{center}
\includegraphics[width=10cm]{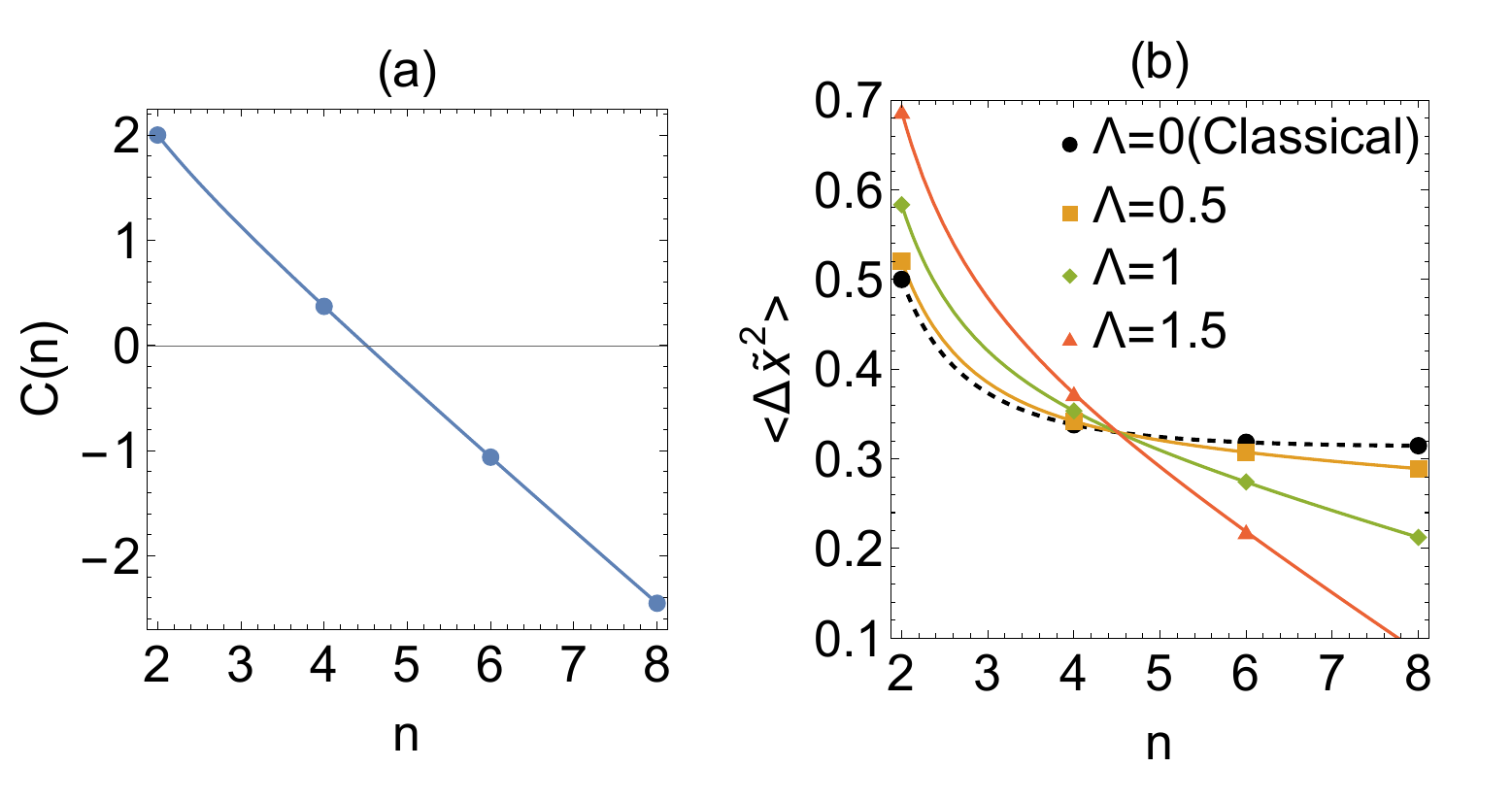} \caption{Results of the
semi-classical approximation. (a) Coefficient of the leading-order
quantum correction, $C(n)$. The coefficient is positive ($C(n) > 0$) for
$n=2$ and $4$, while it becomes negative ($C(n) < 0$) for $n > 4$. Note
that $C(n)$ is independent of the reduced temperature $\tb$. (b)
Positional uncertainty $\ave{\Delta \tx^2}$ for $\tb=1$ as a function of
$\Lambda$ for various values of $n$. For $n=2$ and $4$,
$\ave{\Delta\tx^2}$ increases with increasing $\Lambda$, indicating that
quantum fluctuations enhance the positional uncertainty. In contrast,
for $n>4$, $\ave{\Delta\tx^2}$ decreases with increasing $\Lambda$,
demonstrating the suppression of positional uncertainty by quantum
fluctuations.}  \label{100307_26Nov24}
\end{center}
\end{figure}

For the potential $\tV(\tx)=\tx^n$, we get
\begin{align}
\ave{\Delta\tx^2} \approx \ave{\Delta\tx^2}_{0} 
+\frac{\Lambda^2}{24} C(n),\label{180248_17Jan25}
\end{align}
where
\begin{align}
\ave{\Delta\tx^2}_{0} = \tb^{-2/n}\frac{\Gamma(3/n)}{\Gamma(1/n)},
\end{align}
and 
\begin{align}
C(n)&= 2-\left(\ave{\tV''(R)R^2}_{0}-\ave{\tV''(R)}_{0}
 \ave{R^2}_{0}\right),\new 
&=2-n(n-1)
\left(\frac{\Gamma(\frac{1+n}{n})}{\Gamma(\frac{1}{n})}
-\frac{\Gamma(\frac{3}{n})\Gamma(\frac{n-1}{n})}{\Gamma(\frac{1}{n})^2}
 \right).\label{200051_20Nov24}
\end{align}
As shown in Fig.~\ref{100307_26Nov24}~(a), the coefficient $C(n)$ is
positive for $n=2$ and $4$, but becomes negative for $n>4$. For $n=2$
and $4$, the positive contribution increases $\ave{\Delta\tx^2}$ with
$\Lambda$, indicating that quantum fluctuations enhance the positional
uncertainty, as illustrated in Fig.~\ref{100307_26Nov24}~(b). In
contrast, for $n>4$, quantum fluctuations suppress the positional
uncertainty at small $\Lambda$, as seen in
Fig.~\ref{100307_26Nov24}~(b). Note that this trend is
independent of the temperature for the current model, as $C(n)$ does not
explicitly involve $\tb$. However, for more general potential shapes,
the sign of the leading-order correction may depend on $\tb$,
introducing a possible temperature dependence.

\begin{figure}[t]
\begin{center}
\includegraphics[width=10cm]{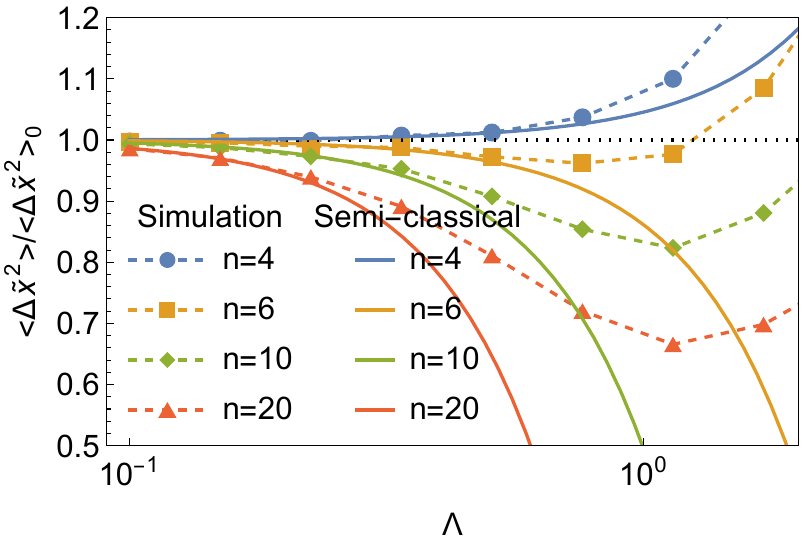} \caption{Comparison between
numerical simulations and semi-classical calculations for
$\tb=1$. Markers represent the same numerical results as in
Fig.~\ref{100335_29Oct24}, while solid lines indicate the results of the
semi-classical approximation. The numerical and theoretical results
agree well for small $\Lambda$, accurately capturing the initial trends
in $\ave{\Delta \tx^2}$. However, the results deviate for large
$\Lambda$, where higher-order corrections become significant.}
\label{225603_3Nov24}
\end{center}
\end{figure}

In Fig.~\ref{225603_3Nov24}, we compare the semi-classical result,
Eq.(\ref{180248_17Jan25}), with the numerical results. The
semi-classical approximation agrees well with the numerical results for
small $\Lambda$, particularly in capturing the initial increase or
decrease of $\ave{\Delta \tx^2}$. However, higher-order corrections
become significant for large $\Lambda$, invalidating the semi-classical
approximation. Notably, the semi-classical calculation fails to
reproduce the non-monotonic behavior of $\ave{\Delta \tx^2}$ observed
for $n > 4$. This is expected, as the semi-classical approximation
considers only the $O(\Lambda^2)$ terms, which can describe only
monotonic increases or decreases in $\ave{\Delta \tx^2}$. Higher-order
corrections are necessary to fully account for the non-monotonic
behavior, which we leave as future work.

\section{Summary and Discussions}
\label{172700_11Nov24}

In summary, we have investigated the effects of quantum fluctuations on
the positional uncertainty $\ave{\Delta x^2}$ of a particle confined in
a non-linear potential, $V(x) = kx^n$. Our results demonstrate that
quantum fluctuations suppress $\ave{\Delta x^2}$ for sufficiently strong
non-linear potentials and small de Broglie wavelengths $\Lambda$. For
larger $\Lambda$, tunneling effects dominate, leading to an increase in
$\ave{\Delta x^2}$ and resulting in a non-monotonic dependence of
$\ave{\Delta x^2}$ on $\Lambda$.

The phenomena observed in our single-particle model may also be relevant
to many-body quantum systems. For instance, previous studies have
reported that quantum fluctuations enhance
crystallization~\cite{hansen1971,runge1988,sese2007computational,sese2007computationalII,sese2013path,yamashita2014gas}
and
vitrification~\cite{markland2011,zamponi2011,markland2012,biroli2012tentative,kinugawa2021,das2021,das2022,winer2024}
for small $\Lambda$. An intuitive explanation for these observations is
that quantum fluctuations effectively increase the particle radius,
thereby reducing the free volume~\cite{markland2011,zamponi2011}. This
mechanism resembles the behavior observed in our model, where quantum
fluctuations reduce $\ave{\Delta x^2}$ for sufficiently strong
non-linear potentials and small $\Lambda$, indicating stronger
confinement of the particle within a narrower region and a corresponding
reduction in free volume~\footnote{Although $\langle \Delta x^2 \rangle$
can be interpreted as the particle radius in some contexts, in this
case, it is more appropriate to consider it as a measure of the free
volume, since $\langle \Delta x^2 \rangle$ grows as the available region
of the particle increases.}. Moreover, the non-monotonic $\Lambda$
dependence of $\ave{\Delta x^2}$ observed in Fig.\ref{100335_29Oct24} is
reminiscent of the non-monotonic behavior of the glass transition point
reported in previous studies\cite{markland2011}. These similarities
suggest that our model may provide a minimal framework for understanding
the non-monotonic behavior observed in studies of glass
transitions~\cite{markland2011,zamponi2011}. However, it is important to
note that our study focuses on a single-particle system, whereas glass
transitions involve many-body interactions. Further studies are needed
to establish a more direct connection between these problems.

In the hard-wall limit ($n \to \infty$), $\ave{\Delta x^2}$
monotonically decreases with increasing $\Lambda$. This behavior
contrasts with predictions from mode-coupling theory (MCT), a dynamical
mean-field theory of glass
transitions~\cite{gotze1999recent,reichman2005mode,gotze2009complex},
which suggests a non-monotonic dependence of the glass transition point
on $\Lambda$ even for hard-sphere
potentials~\cite{markland2011,markland2012}. Recently, an exact
calculation for quantum hard spheres has been performed in the limit of
large spatial dimensions~\cite{winer2024}, using the replica liquid
theory (RLT), a static mean-field theory of glass
transitions~\cite{edwards1975,mezard1987spin,mezard1999first,charbonneau2017glass,parisi2010,parisi2020theory}.
The RLT predicts a monotonic decrease of the glass transition point with
increasing $\Lambda$, which is qualitatively consistent with our
findings in the hard-wall limit. Additionally, studies on the
crystallization of quantum hard spheres have consistently reported a
monotonic decrease in the transition point with increasing
$\Lambda$~\cite{hansen1971,runge1988,sese2007computational,sese2007computationalII,sese2013path}. This
consistency leads us to speculate that the non-monotonic behavior
predicted by MCT for quantum hard spheres may be an artifact of the
approximation. Further investigations would be valuable to clarify this
point.

In previous studies of the glass transition, non-monotonic behaviors are
reported for the ratio $\langle u_i^2 \rangle / \langle
u_i^2\rangle_{\rm free}$, which quantifies local fluctuations around the
center of
mass~\cite{markland2011,markland2012,kinugawa2021,tsujimoto2024}. In
contrast, our preliminary numerical results show that this ratio
decreases monotonically with $\Lambda$ in our model. We attribute this
difference to the simplicity of the potential used here, which has a
single minimum. Even at large $\Lambda$, tunneling remains limited, and
$\langle u_i^2 \rangle$ continues to be governed by the single-well
confinement. Introducing a more complex landscape with multiple minima,
as considered in glassy systems, may recover the non-monotonic
behavior~\cite{phillips1972,stillinger1995,sastry1998,debenedetti2001supercooled}. We
leave a systematic investigation of this possibility for future work.

\ack  
We thank A.~Ikeda, M.~Udagawa, and K.~Miyazaki, for
helpful discussions. I acknowledge the use of OpenAI’s ChatGPT
(https://chat.openai.com/) for assistance in improving the clarity,
grammar, and overall readability of the manuscript. This work was
supported by KAKENHI 23K13031.

\section*{References}
\bibliographystyle{iopart-num.bst}
\bibliography{reference}

\end{document}